\documentclass[twocolumn,showpacs,superscriptaddress,amsmath,amssymb]{appolb}
\usepackage{graphicx}
\usepackage{dcolumn}
\usepackage{bm}

\def\bd{
\begin{document}} \def\ed{\end{document}}
\def\bmp{\begin{minipage}} \def\emp{\end{minipage}}
\def\bcc{\begin{center}} \def\ecc{\end{center}}     \def\npg{\newpage}
\def\beq{\begin{equation}} \def\eeq{\end{equation}} \def\hph{\hphantom}
\def\be{\begin{equation}} \def\ee{\end{equation}} \def\r#1{$^{[#1]}$}
\def\n{\noindent} \def\ni{\noindent} \def\pa{\parindent}
\def\hs{\hskip} \def\vs{\vskip} \def\hf{\hfill} \def\ej{\vfill\eject}
\def\cl{\centerline} \def\ob{\obeylines}  \def\ls{\leftskip}
\def\underbar#1{$\setbox0=\hbox{#1} \dp0=1.5pt \mathsurround=0pt
   \underline{\box0}$}   \def\ub{\underbar}    \def\ul{\underline}
\def\f{\left} \def\g{\right} \def\e{{\rm e}} \def\o{\over} \def\d{{\rm d}}
\def\vf{\varphi} \def\pl{\partial} \def\cov{{\rm cov}} \def\ch{{\rm ch}}
\def\la{\langle} \def\ra{\rangle} \def\EE{e$^+$e$^-$} \def\pt{p_{\rm t}}
\def\bitz{\begin{itemize}} \def\eitz{\end{itemize}}
\def\btbl{\begin{tabular}} \def\etbl{\end{tabular}}
\def\btbb{\begin{tabbing}} \def\etbb{\end{tabbing}}
\def\beqar{\begin{eqnarray}} \def\eeqar{\end{eqnarray}}
\def\\{\hfill\break} \def\dit{\item{-}} \def\i{\item}
\def\bbb{\begin{thebibliography}{9} \itemsep=-1mm}
\def\ebb{\end{thebibliography}} \def\bb{\bibitem}
\def\bpic{\begin{picture}(260,240)} \def\epic{\end{picture}}
\def\akgt{\cl{\bf ACKNOWLEDGMENTS}}
\def\fgn{\noindent{\bf\large\bf figure captions}}
\def\lan{\langle}
\def\ran{\rangle}
\def\p{\pi}
\def\ifmath#1{\relax\ifmmode #1\else $#1$\fi}%
\def\rc{\ifmath{{\mathrm{c}}}}
\def\cut{\ifmath{{\mathrm{cut}}}}
\def\rF{\ifmath{{\mathrm{F}}}}
\def\rK{\ifmath{{\mathrm{K}}}}
\def\rp{\ifmath{{\mathrm{p}}}}
\def\rt{\ifmath{{\mathrm{t}}}}
\def\LAB{\ifmath{{\mathrm{LAB}}}}
\def\cut{\ifmath{{\mathrm{cut}}}}
\def\beq{\begin{equation}}
\def\eeq{\end{equation}}

\newcommand{\cinst}[2]{$^{\mathrm{#1}}$~#2\par}
\newcommand{\crefi}[1]{$^{\mathrm{#1}}$}
\newcommand{\crefii}[2]{$^{\mathrm{#1,#2}}$}
\newcommand{\crefiii}[3]{$^{\mathrm{#1,#2,#3}}$}
\newcommand{\HRule}{\rule{0.5\linewidth}{0.5mm}}

\bd

\title{Entropy Analysis in $\pi^{+}\rp$ and $\rK^{+}\rp$ Collisions at 
$\sqrt{s}=22$ GeV}

\author{M.R.~Atayan$^1$, Bai Yuting$^{2,a}$, E.A.~De Wolf$^3$, 
A.M.F.~Endler$^4$, Fu Jinghua$^2$, H.~Gulkanyan$^5$, R.~Hakobyan$^{5,b}$,
W.~Kittel$^6$, Liu Lianshou$^2$, Li Zhiming$^{2,6}$, Z.V.~Metreveli$^{7,c}$,
W.J.~Metzger$^6$, L.N.~Smirnova$^8$, L.A.~Tikhonova$^8$, 
A.G.~Tomaradze$^{7,c}$, Wu Yuanfang$^2$, S.A.~Zotkin$^{8,d}$
\address{
$^1$ Institute of Physics, AM-375036 Yerevan, Armenia\\
$^2$ Institute of Particle Physics, Hua-Zhong Normal
University, Wuhan 430079, China; \\
$^3$ Department of Physics, University of Antwerp, B-2610
Wilrijk, Belgium\\
$^4$ Centro Brasileiro de Pesquisas Fisicas, BR-22290 Rio de
Janeiro, Brazil\\
$^5$ Institute of Physics, AM-375036 Yerevan, Armenia\\
$^6$ Radboud University/NIKHEF, NL-6525~ED Nijmegen, The
Netherlands\\
$^7$ Institute for High Energy Physics of Tbilisi State
University, GE-380086 Tbilisi, Georgia\\
$^8$ Scobeltsyn Institute of Nuclear Physics, Lomonosow
Moscow State University, RU-119899 Moscow, Russia\\
$^a$ Now at University of Utrecht/NIKHEF, The Netherlands\\
$^b$ Now at University of Regina, Saskatchewan, S4S 4G2,
Canada\\
$^c$ Now at Northwestern Univ., Evanston, U.S.A.\\
$^d$ Now at DESY, Hamburg, Germany\\ ~\\
EHS/NA22 Collaboration}}
\maketitle
\begin{abstract}
The entropy properties are analyzed by Ma's coincidence method in
$\pi^{+}\rp$ and $\rK^{+}\rp$ collisions of the NA22 experiment at
250 GeV/$c$ incident momentum. By using the R\'{e}nyi entropies,
we test the scaling law and additivity properties in rapidity
space. The behavior of the R\'{e}nyi entropies as a function of
the average number of particles is investigated. The results are
compared with those from the {\sc Pythia} Monte Carlo event
generator.
\end{abstract}


\section{Introduction}

It was suggested recently that the event coincidence probability method of measuring
entropy, originally proposed by Ma~\cite{Ma}, is well suited for the analysis of local
properties in multiparticle systems produced in high energy
collisions~\cite{Bialas1,Bialas2,Bialas3, Fialkowski,Bialas4,Bialas5,Bialas6}. In heavy
ion collisions it can be used in the search for formation of a quark gluon plasma
(QGP)~\cite{Bialas1}, but the method is equally applicable to hadron-hadron
collisions~\cite{Fialkowski}. It  also proved effective for the study of systems of
particles created in multiplicative branching processes~\cite{Bialas6} as present in
high-energy QCD jet fragmentation. Finally, entropy measurements offer an additional
tool to analyze event-to-event fluctuations and particle
correlations~\cite{Bialas3,Bialas4}.

The existence of a QGP at high energy density is a prediction of QCD. Many theorists and
experimentalists are searching for possible signals of the formation of such a new state
of matter. The primary goal of the Relativistic Heavy Ion Collider (RHIC) at Brookhaven
National Laboratory is to create and to study this deconfined state. Systematic
measurements of the local entropy at RHIC may provide direct information about the
internal degrees of freedom of the QGP state and its evolution. It is thus very
important to first see how the proposed entropy measures behave in hadron-hadron
collisions.

In this note, we present the results of a study of entropy in
$\pi^{+}\rp$ and $\rK^{+}\rp$ collisions at 250 GeV/$c$ incident
momentum ($\sqrt{s}=22$ GeV) collected by the NA22 experiment. We
investigate the dependence on discretization of the system, the
effects of particle correlations by testing scaling and additivity
properties of the entropy measures, as well as the multiplicity
dependence of the R\'{e}nyi entropies in rapidity space. We also
compare the results of the NA22 data to those of the Monte Carlo
event generator {\sc Pythia} 5.7~\cite{pythia} used for inelastic,
non-single-diffractive pion-proton collisions with Bose-Einstein
Correlations (BEC) included, as well as to a simple random
production model. We end with a short conclusion and outlook.

\section{\bf Procedure and variables}
The original definitions of the standard Shannon entropy and the
R\'{e}nyi entropies are:
\beqar  
S = - \sum _j p _j \ln p _j \hskip 25bp H_{k} = \sum_j (p _j)^{k}.
\eeqar

\noindent Here, $p _j = n _j/N$ denotes the probability to obtain
a specific configuration of the system, where $n _j$ is the number
of events in such a configuration and $N$ is the total number of
events. The sum runs over all possible configurations.

Following Ma's method~\cite{Ma} and as explained
in~\cite{Bialas3}, the first step in the entropy measurement is to
determine coincidence probabilities. For every event, a certain
phase space region is divided into $M$ bins of equal size. An
event is then characterized by the number of particles, $m_i$, in
each bin, i.e., by a set of integer numbers $s\equiv \{m_{i}\}$,
where $i=1, ...M$.

After counting how many times, $n_{s}$,  the set $s$ appears in the whole event sample,
one can determine the sums
\beqar  
N_k= \sum_s n_s(n_s-1)...(n_s-k+1)    .
\eeqar

\noindent This gives the total numbers of observed coincidences of
$k$ configurations.

The coincidence probability of $k$ configurations is then given by~\cite{Ma}
\beqar  
C_k = \frac{N_k}{N(N-1)...(N-k+1)}   \;\;\;,
\eeqar

\noindent where $N$, as above, is the total number of events in
the sample. Only states with $n_{s}\geq k$ contribute to $C_{k}$.

From the coincidence probabilities, one calculates the R\'{e}nyi
entropies defined as~\cite{Renyi}
\beqar  
H_k\equiv - \frac{\ln C_k}{k-1}     .
\eeqar

The Shannon entropy $S$ is formally equal to the limit of $H_{k}$ as
$k\rightarrow 1$. So it can only be obtained by an extrapolation method. One
possibility we will investigate in some detail is to take~\cite{Bialas2}
\beqar  
H_k= a\frac {\ln k}{k-1} + a_0+ a_1(k-1) +a_2(k-1)^2 +....\;\;\;.
\eeqar

These expansions  may be used as a system of linear equations to
determine $a$ and $a_i$, with as many terms as the number of
$H_k$'s available. The Shannon entropy is then calculated as
$S=a+a_0$. It is instructive to see if the value of S depends on
the number of R\'{e}nyi entropies used. In general, it turns out
that already the value obtained from two terms, $S[H_{2},H_{3}]$,
and that from three terms, $S[H_{2},H_{3},H_{4}]$, nearly
coincide.

One of the most attractive features of Ma's coincidence method
compared to the standard method of using Eq.~(1) directly is that,
as seen from Eq.~(3), the statistical error decreases very fast
with increasing number of configurations. Moreover, it has been
demonstrated~\cite{Fialkowski} that the extrapolation method
yields more stable results than the standard method.

For a system close to thermal equilibrium, and if the phase space subdivision is
sufficiently fine-grained, the scaling relation
\beqar  
H_{k}(lM)=H_{k}(M)+\ln l \Rightarrow S(lM)=S(M)+\ln l \eeqar
\noindent holds. Here, $M$ and $lM$ are the numbers of bins in two
different discretizations.

Another feature is additivity: the entropies measured in a region $R$ which is the union
of two non-overlapping and independent regions $R_{1}$ and $R_{2}$ satisfy
\beqar  
H_{k}(R) & = & H_{k}(R_{1})+H_{k}(R_{2}) \nonumber\\
         & \Rightarrow &
     S(R)=S(R_{1})+S(R_{2})             .
\eeqar

\noindent We will check these two important properties below.

\section{Data sample}
The NA22 experiment made use of the European Hybrid Spectrometer (EHS)
in combination with the Rapid Cycling Bubble Chamber (RCBC).
Details on the setup and on the reconstruction
of the data can be found in~\cite{VHTH87,DRTH88}.

Charged-particle momenta are measured over the full solid angle with an
average resolution varying from 1-2\% for tracks reconstructed in RCBC and
1-2.5\% for tracks reconstructed in the first lever arm, to 1.5\% for tracks
reconstructed in the full spectrometer. Ionization information is used to
identify and exclude
protons up to 1.2 GeV/$c$ and electrons (positrons) up to 200 MeV/$c$.
All unidentified tracks are given the pion mass.

In our analysis, events are accepted when the measured and reconstructed
charged-particle multiplicity are the same, no electron is detected among the secondary
tracks and the number of badly reconstructed tracks is 0. The selection of the data is
described in detail in~\cite{SelectEvent}. After all necessary rejections, a total of
44,524 inelastic, non-single-diffractive events is obtained. Acceptance losses are
corrected by a multiplicity-dependent event weighting procedure.

\section{Results and Discussions}

\begin{figure}
\centering
\includegraphics[width=10.0cm]{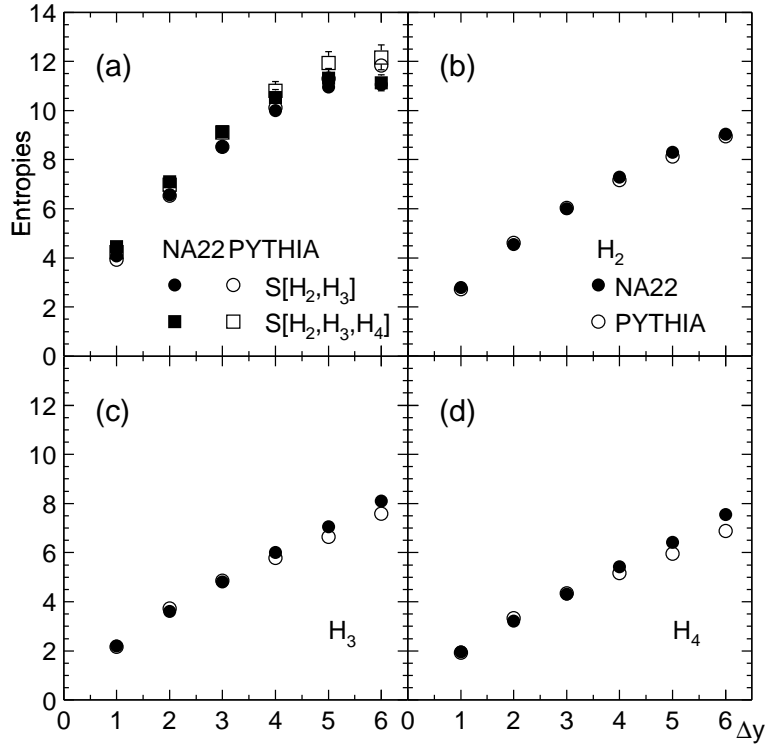}
\caption{\label{Fig. 1} Shannon and R\'{e}nyi entropies obtained
from an $M=9$-fold division of central rapidity windows of size
$\Delta y$ ranging from $1$ to $6$ units. }
\end{figure}

In Fig.~1, we show the Shannon entropy and the R\'{e}nyi entropies 
$H_{2}$ to $H_{4}$ as obtained from an $M=9$-fold division of
central cms rapidity windows of size $\Delta y$ ranging from $1$
to $6$ units. The Shannon entropy is calculated by extrapolation
according to Eq.~(5).
The solid symbols show the NA22 data, the open symbols the {\sc
Pythia} predictions. All R\'{e}nyi and Shannon entropies increase
as the rapidity window widens. The values of $H_k$ decrease with
increasing $k$. The figure further shows that  the two-term
extrapolated Shannon entropy $S[H_{2},H_{3}]$ agrees well with the
three-term extrapolation $S[H_{2},H_{3},H_{4}]$. For the $H_k$,
{\sc Pythia} tends to be flatter than the data. This trend
increases with increasing order $k$, so that the extrapolation
procedure leads to a Shannon entropy $S$ which rises more steeply
than the data.

It may be worthwhile to point out that, in general, the
extrapolation procedure for the Shannon entropy is not unique and
thus will introduce an additional uncertainty~\cite{Bialas2,Bialas3}.
For example, using a polynomial expansion (Eq.~(20) of ref. [3]) for
$\Delta y = 6$ results in a value which is about 10\% smaller for
$S[H_{2},H_{3}]$ and 7\% smaller for $S[H_{2},H_{3},H_{4}]$.
 The
R\'{e}nyi entropies are of great interest by themselves and
provide valuable information about the system without the need for
any extrapolation. In the following, we therefore concentrate on
the R\'{e}nyi entropies, in particular on $H_{2}$.

If scaling according to Eq.~(6) holds, one should observe a linear relation if
$H_{k}(M)$ is plotted as a function of $-\ln\delta y$, where $\delta y = {\Delta y}/{M}$
is the bin size.

\begin{figure}
\centering
\includegraphics[width=8cm]{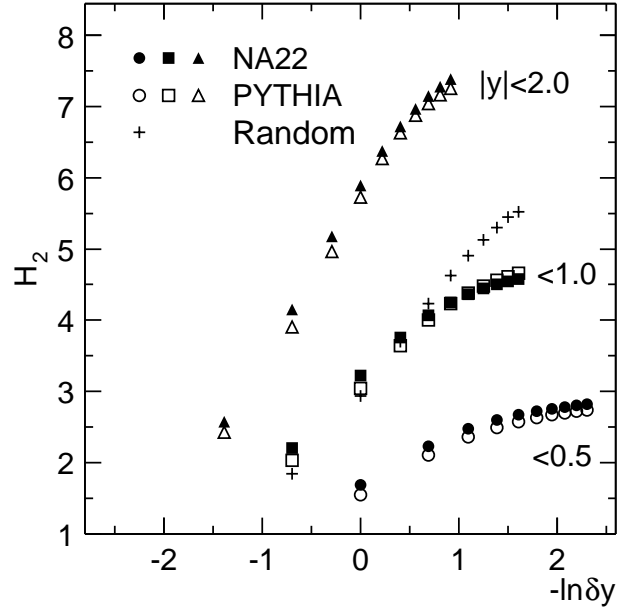}
\caption{\label{Fig. 2}
The dependence on the dividing bin sizes of the second R\'{e}nyi entropy
calculated in central rapidity windows.
}
\end{figure}

We first test scaling of $H_{2}$ in central rapidity windows in
Fig.~2. The solid circles, squares and triangles represent the
NA22 data for $|y|<0.5, 1.0, 2.0$, respectively. They all are
flattening with increasing $-\ln\delta y$, i.e., the scaling law
does not hold for the NA22 data. This implies that
there are strong particle correlations in the process under
investigation. The open symbols in Fig.~2 represent the {\sc
Pythia} results with a Bose-Einstein correlation strength
$\lambda=1.0$~\cite{pythia}. They do not fully agree with the
data. We have verified that this discrepancy is not sensitive to
the actual value of $\lambda$ used.

The cross symbols for $|y|<1.0$ in Fig.~2 show the results from a random model generated
as follows: take the multiplicity distribution in the window $|y|<1.0$ from the data,
but the multiplicity fluctuations in each $\delta y$ bin to be Poissonian with a mean as
in the data. In this model, there are no correlations between particles, so it should
give a nearly straight-line relationship, as indeed shown by the figure.

\begin{figure}
\centering
\includegraphics[width=12.2cm]{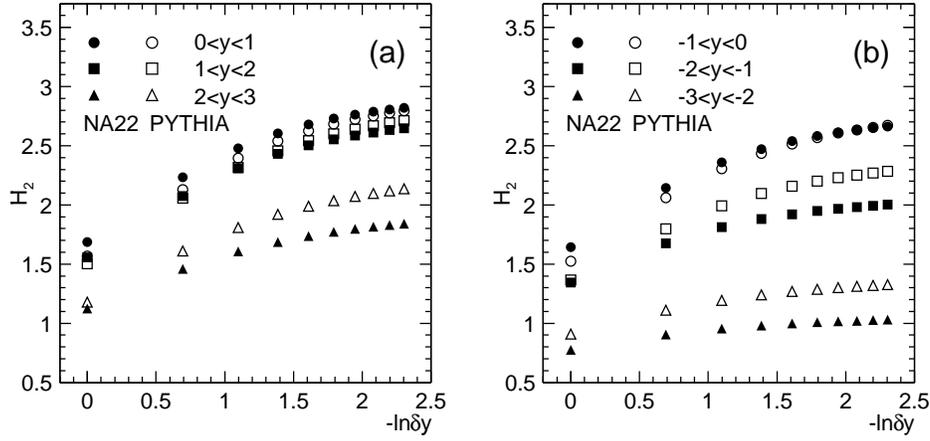}
\caption{\label{Fig. 3}
The dependence on the dividing bin sizes of the second R\'{e}nyi entropy
calculated in (a) forward and (b) backward hemispheres in different
non-central rapidity windows.
}
\end{figure}

Due to the asymmetry of the proton and meson fragmentation regions
in the NA22 experiment, we plot forward and backward hemispheres
separately in different non-central windows of fixed size $\Delta
y=1$ in Fig.~3 (a) and (b), respectively. We see that, again, the
$H_{2}$ values are flattening with increasing $-\ln\delta y$, so
that no linear relationship is found. {\sc Pythia} has the same
trend, but overestimates the data in the peripheral regions.
We also note that
$H_2$ becomes smaller for more peripheral windows, due to
decreasing particle density, i.e., decreasing coincidence
probability.

\begin{figure}
\vskip 3mm
\centering
\includegraphics[width=12.2cm]{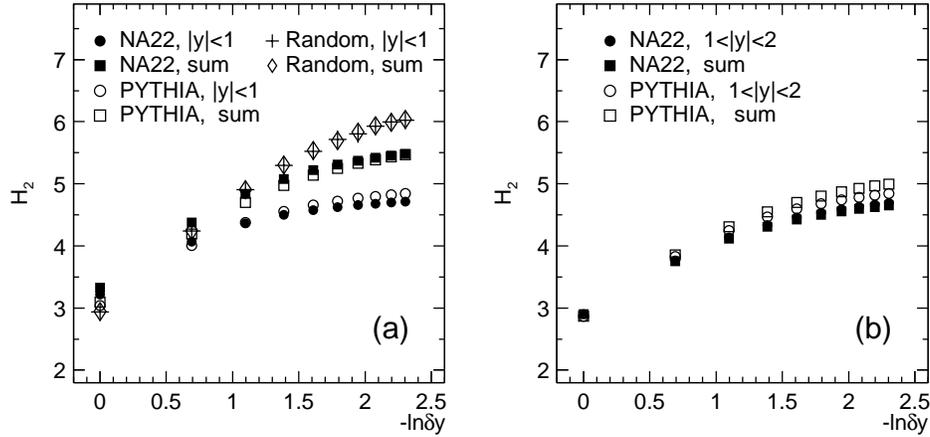}
\caption{\label{Fig. 4}
The second R\'{e}nyi entropy as a function of $-\ln\delta y$ measured
in two (a) adjacent and (b) separate intervals.
}
\end{figure}

In order to test additivity, Fig.~4~(a) shows $H_{2}$ obtained in
two ways: one is a direct measurement in the rapidity window
$|y|<1$ (solid circles), the other the sum of the values obtained
from the adjacent ranges $-1<y<0$ and  $0<y<1$ (solid squares). A
clear difference is observed between these two results. The
$H_{2}$ obtained from summation is larger than the one calculated
directly, and the difference increases with increasing $-\ln\delta
y$. So, additivity of $H_2$ is not observed here. The reason is
that strong correlations exist between the particles belonging to
these two adjacent $y$ ranges. Indeed, no  violation of additivity
is observed for the random model (crosses and diamonds in
Fig.~4~(a)), thus confirming the role of correlations in the data.

Fig.~4 (b) shows the results from two regions separated by two
rapidity units. In this case, the difference between the direct
measurement in the window $1<|y|<2$ (solid circles) and the sum of
the two non-adjacent windows $-2<y<-1$ and $1<y<2$ (solid squares)
is very small. {\sc Pythia} (open symbols) shows the same trend
but again overestimates the values in the non-central rapidity
regions.

The additivity observed here is consistent with the experimental
observation that correlations between particles belonging to
widely separated rapidity intervals are small~\cite{na22corr}.
However, at much higher energies, strong long-range correlations
are observed~\cite{ua5corr} which should lead to a breaking of the
additivity property. This was confirmed in {\sc Pythia} model
calculations for proton-proton collision at $\sqrt{s}=1800$
GeV~\cite{Fialkowski}.

As advocated in~\cite{Bialas4}, the dependence of R\'{e}nyi
entropies on particle multiplicity $n$ carries important
information on the produced system. An entropy proportional to $n$
is indicative of an equilibrated system with no strong long-range
correlations. On the other hand, for non-equilibrated systems or a
superposition of sub-systems with different properties,
proportionality to $\ln n$ is expected.

\begin{figure}
\centering
\includegraphics[width=12.2cm]{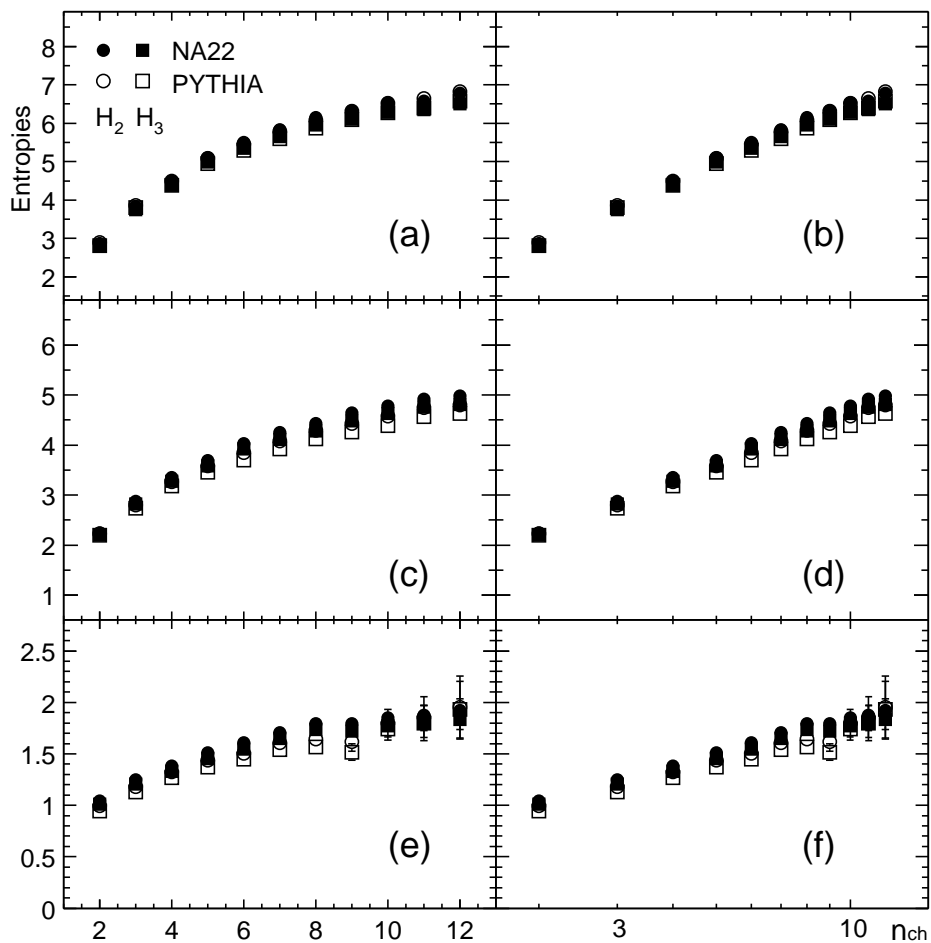}
\caption{\label{Fig. 5} The dependence of the second and third
R\'{e}nyi entropies on multiplicity in linear (left) and
logarithmic (right) scale, with (a)(b) $M=6$-fold division in the
rapidity window $|y|<3$, (c)(d) $M=4$ in $|y|<2$ and (e)(f) $M=2$
in $|y|<1$.
}
\end{figure}

\vskip0.3cm
\begin{table}
\caption {Fit values of R\'{e}nyi entropies in different rapidity
windows.}
{\begin{center} {
\begin{tabular}{ccccc}
\hline
\hline
&   &  $c_1$ & $c_2$ &  $\chi^2$/NDF  \\
\hline
$|y|<3, M=6$ & $H_2$ & 1.54$\pm$0.06 & 2.18$\pm$0.04 &46.23/8  \\
& $H_3$ &1.60$\pm$0.06 & 2.06$\pm$0.04 &49.15/8  \\
$|y|<2, M=4$ & $H_2$ & 1.17$\pm$0.06 & 1.57$\pm$0.03 &  5.77/8  \\
& $H_3$ &1.22$\pm$0.05 & 1.48$\pm$0.03 &  7.14/8  \\
$|y|<1, M=2$ & $H_2$ & 0.68$\pm$0.06 & 0.52$\pm$0.04 &  1.79/8  \\
& $H_3$ &0.65$\pm$0.05 & 0.50$\pm$0.03 &4.22/8  \\
\hline \hline
\end{tabular}
}
\end{center}
}
\end{table}

In Fig.~5, we show the charged-particle multiplicity dependence of
$H_{2}$ (solid circles) and $H_{3}$ (solid squares) for three
different rapidity windows, where the number of divisions $M$ is
chosen such that $\delta y$ is the same for each rapidity window.
Table I shows the fit results for the logarithmic function
\beqar  
y=c_{1}+c_{2}\ln n_{\rm{ch}} . \eeqar In the central region, a
logarithmic rather than a linear relation is observed. This again
confirms that there is no thermal equilibrium in the system. {\sc
Pythia} (open symbols) agrees quite well with the data. \\

\section{Conclusions and outlook}
We have analyzed the entropy properties of multiparticle
production in $\pi^{+}\rp$ and $\rK^{+}\rp$ collisions at 250
GeV/$c$. By using R\'{e}nyi entropies, we find that both the
scaling and the additivity properties are not generally valid and
that the multiplicity dependence shows a logarithmic rather than a
linear relation. All of these confirm the expectation that thermal
equilibrium is not reached in hadron-hadron collisions at
$\sqrt{s}=22$ GeV.

The {\sc Pythia} Monte Carlo model agrees, in general, quite well
with the data. However, significant deviations exist. In
particular, the model overestimates the values in the peripheral
rapidity regions, presumably due to too weak correlations. This
shows that R\'{e}nyi entropies provide a new sensitive measure of
multiparticle correlations.

It would be interesting to investigate the entropy properties of
high-temperature, high-density systems, which may create the long
expected QGP. RHIC has already collected data on gold-gold
collisions at 130 and 200 GeV and pp collisions at 200 GeV. The
results presented here should provide a valuable guide to the
interpretation of these measurements.

\vskip 0.5cm
\akgt
\vskip 0.2cm

We thank A. Bia{\l}as for constructive discussions and remarks. We
are grateful to K. Fia{\l}kowski and R. Wit for the warmhearted
discussions concerning the {\sc Pythia} model. One of the authors,
Li Z, would like to thank Nuffic, The Netherlands, for financial
support. This work is part of the research program of the
``Stichting voor Fundamenteel Onderzoek der Materie (FOM)", which
is financially supported by the ``Nederlandse Organisatie voor
Wetenschappelijk Onderzoek (NWO)". We further thank NWO for
support within the program for subsistence to the former Soviet
Union (07-13-038). The Yerevan group is financially supported, in
the framework of the theme No. 0248, by the Government of the
Republic of Armenia. This work is also supported in part by the
National Natural Science Foundation of China under Grant 10475030
and the Ministry of Education of China and by the Royal Dutch
Academy of Sciences under the Project numbers 01CDP017, 02CDP011
and 02CDP032 and by the U.S. Department of Energy.

\ed